\documentclass[reprint,aps,pra]{revtex4-2}
\usepackage{graphicx} \usepackage{amsmath,amssymb}  \usepackage{bm}
\usepackage{color}

\begin{document}

\title{
    1/$f$ noise of a tiny tunnel magnetoresistance sensor originated from a wide distribution of bath correlation time
}

\author{Hiroshi Imamura}\email{h-imamura@aist.go.jp}
\author{Hiroko Arai}\email{arai-h@aist.go.jp}
\author{Rie Matsumoto}
\author{Toshiki Yamaji}
\affiliation{ National Institute of Advanced Industrial Science and
    Technology (AIST), Research Center for Emerging Computing Technologies (RCECT), Tsukuba, Ibaraki
    305-8568, Japan }

\begin{abstract}
    Tunnel magnetoresistance (TMR) sensor is a highly sensitive magnetic field sensor and is expected to be applied in various fields, such as magnetic recording, industrial sensing, and bio-medical sensing. To improve the detection capability of TMR sensors in low frequency regime it is necessary to suppress the 1/$f$ noise. We theoretically study 1/$f$ noise of a tiny TMR sensor using the macrospin model. Starting from the generalized Langevin equation, 1/$f$ noise power spectrum and the Hooge parameter are derived. The calculated Hooge parameter of a tiny TMR sensor is much smaller than that of a conventional TMR sensor with large junction area. The results provide a new perspective on magnetic 1/$f$ noise and will be useful for improvement of TMR sensors.
\end{abstract}

\maketitle

\section{INTRODUCTION}
Tunnel magnetoresistance (TMR) sensor \cite{Freitas2007,Egelhoff2009,Lei2011,Fujiwara2018,Wang2019,Oogane2021,Nakatani2022} is a highly sensitive magnetic field sensor where the magnetic field signal is converted to the change in resistance of a magnetic tunnel junction (MTJ) \cite{Julliere1975,Maekawa1982,Miyazaki1995,Parkin2004,Yuasa2004}. The most popular application of the TMR sensor is a reading head of hard disk drives. Because of its high sensitivity, small size, and low power consumption, the TMR sensors are expanding their applications into a variety of fields such as industrial sensing and bio-medical sensing. In the bio-medical applications such as magetocardiography and magnetoencephalography, TMR sensors detect the weak magnetic fields generated in the human heart and brain by electrophysiological activity of cardiac muscle and nerve cells \cite{Fujiwara2018,Wang2019,Oogane2021}. The frequency range of the bio-magnetic signal is less than a few hundred Hz where the 1/$f$ noise is the dominant noise. Reduction of 1/$f$ noise is a key issue for bio-medical applications \cite{Pannetier2005,Lei2011}.

1/$f$ noise is a ubiquitous low-frequency noise whose noise power is inversely proportional to the frequency, $f$ \cite{McWhorter1957,Hooge1981,Dutta1981,Weissman1988}. A large number of theories have been developed to explain the mechanism of 1/$f$ noise as reviewed in Ref. \cite{Weissman1988}. An obvious way to obtain a 1/$f$ power spectrum is to superimpose a large number of Lorentzian power spectra produced by exponential relaxation processes \cite{Bernamont1937,duPre1950,VanDerZiel1950,McWhorter1957,Weissman1988}. The magnitude of the 1/$f$ noise in different devices and materials is characterized by the Hooge parameter \cite{Hooge1981}.

The magnetic 1/$f$ noise derived from thermal fluctuation of magnetization in a TMR sensor has been studied by several groups \cite{Ingvarsson2000,Ren2004,Jiang2004,Gokce2006,Almeida2006,Egelhoff2009,Silva2015,Garcia2021,Nakatani2022,Matos2023}. In most previous studies the TMR sensors exhibit clear hysteresis in the magnetic field dependence of resistance, and the 1/$f$ noise is observed within the hysteresis loop. The observed 1/$f$ noise has been attributed to thermally excited hopping of magnetic domain walls between pinning sites. It is natural to ask the question if the magnetic 1/$f$ noise appears in a tiny TMR sensor where the domain wall cannot be created. If 1/$f$ noise appears in a tiny TMR sensor, what is its power? To answer this question it is necessary to develop a theoretical model of magnetic 1/$f$ noise based on the macrospin model.

In this paper, we propose a theoretical model for the magnetic 1/$f$ noise of a tiny TMR sensor based on the macrospin model. Starting from the generalized Langevin equation, we derive an analytical expression of the voltage power spectrum in the low frequency regime.  Assuming a wide distribution of bath correlation times, the derived voltage power spectrum is inversely proportional to the frequency, i.e. 1/$f$ noise. We also show that the Hooge parameter of a tiny TMR sensor is much smaller than that of a conventional TMR sensors with large junction area.

\section{THEORETICAL MODEL}
\begin{figure}[t]
    \centerline{ \includegraphics[width=\columnwidth]{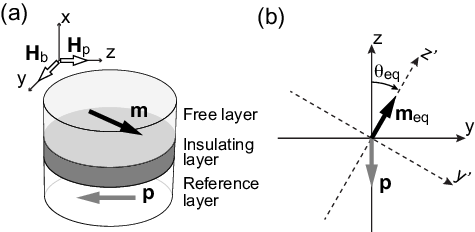}}
    \caption{
        \label{fig:fig1}
        (a) Schematic illustration of a magnetic tunnel junction. The insulating layer is sandwiched by the free layer (FL) and the reference layer (RL). The direction of the magnetization in the FL is denoted by the magnetization unit vector, $\bm{m}$. The direction of the magnetization in the RL is denoted by the magnetization unit vector, $\bm{p}$, and is fixed in the negative $z$ direction. The magnetization in the FL is pinned by the orange peel coupling field, $\bm{H}_{p}$, directing in the $z$ direction and by the uniaxial anisotropy field along the $z$ axis.} The bias field, $\bm{H}_{b}$, is applied in the $y$ direction.
    (b) Definition of the rotated coordinate system. The $z'$ axis is aligned to the equilibrium direction of the magnetization in the FL, $\bm{m}_{\mathrm{eq}}$, by rotating around the $x$ axis with the angle $\theta_{\mathrm{eq}}$.
\end{figure}

The system we consider is the MTJ nano-pillar shown in Fig. \ref{fig:fig1} (a), which is the core element of a tiny TMR sensor. The nonmagnetic insulating layer is sandwiched by the ferromagnetic layers. The top ferromagnetic layer is the free layer (FL) of which magnetization is softly pinned by the orange peel coupling field, $\bm{H}_{p}$, directing in the $z$ direction and by the uniaxial anisotropy field, $H_{k}$, along the $z$ axis. The direction of the magnetization in the FL is denoted by $\bm{m}$. To tune the sensitivity, the bias field, $\bm{H}_{b}$, is applied in the $y$ direction. The bottom ferromagnetic layer is the reference layer of which magnetization unit vector, $\bm{p}$, is fixed to the negative $z$ direction \cite{Nakatani2022}.  The size of the TMR sensor is assumed to be so small that a domain wall cannot be created in the FL, i.e. about or less than 10 nm.

Assuming that the FL is a thin circular disk, the magnetic free energy density of the FL is given by
\begin{align}
    E
     & =
    -\mu_{0} M_{s} \bm{m}\cdot\left(\bm{H}_{p} + \bm{H}_{b}\right)
    + \frac{1}{2}\mu_{0} M_{s}^{2} m_{x}^{2} \nonumber           \\
     & \hspace{1em} -  \frac{1}{2}\mu_{0} M_{s} H_{k} m_{z}^{2},
\end{align}
where $\mu_{0}$ is the permeability of vacuum, and $M_{s}$ is the saturation magnetization. The equilibrium direction, $\bm{m}_{\mathrm{eq}} = (0,\sin\theta_{\mathrm{eq}},\cos\theta_{\mathrm{eq}})$, is obtained by minimizing $E$.

The voltage noise of the TMR sensor is induced by the resistance variation due to fluctuation of $\bm{m}$ around the equilibrium direction. To calculate the fluctuation of $\bm{m}$ we introduce the rotated coordinate system shown in Fig. \ref{fig:fig1}(b), where the $y'$ and $z'$ axes are generated by rotating the $y$ and $z$ axes around the $x$ axis by the angle $\theta_{\mathrm{eq}}$.
The basis vectors of the $x-y'-z'$ coordinate system are defined as
\begin{align}
    \label{eq:basis_transform}
    \begin{pmatrix}
        \bm{e}_{x}  \\
        \bm{e}_{y'} \\
        \bm{e}_{z'}
    \end{pmatrix}
    =
    \begin{pmatrix}
        1 & 0                        & 0                         \\
        0 & \cos\theta_{\mathrm{eq}} & -\sin\theta_{\mathrm{eq}} \\
        0 & \sin\theta_{\mathrm{eq}} & \cos\theta_{\mathrm{eq}}
    \end{pmatrix}
    \begin{pmatrix}
        \bm{e}_{x} \\
        \bm{e}_{y} \\
        \bm{e}_{z}
    \end{pmatrix}.
\end{align}
In the rotated coordinate system, the magnetization unit vector in the FL is represented as
\begin{align}
    \label{eq:m_in_prime_coordinate}
    \bm{m} & = m_{x} \bm{e}_{x} + m_{y'} \bm{e}_{y'} + m_{z'} \bm{e}_{z'}.
\end{align}
Since we are interested in the small fluctuation of $\bm{m}$ around $\bm{m}_{\mathrm{eq}}$, we assume $|m_{x}| \ll 1$, $|m_{y'}| \ll 1$, and $|m_{z'}|\simeq 1$.

The resistance of the TMR sensor is given by \cite{Julliere1975,Maekawa1982,Inoue1996}
\begin{align}
    \label{eq:R_def}
    R = R_{0} + \frac{\bar{R}}{1+P^{2}\bm{m}\cdot\bm{p}},
\end{align}
where $R_{0}$ is the resistance not caused by tunneling, $\bar{R}$ is the resistance due to tunneling at $\bm{m}\cdot\bm{p}=0$, $P$ is the spin polarization of tunneling electrons. Substituting Eq. \eqref{eq:m_in_prime_coordinate} into Eq. \eqref{eq:R_def}, the resistance is obtained as
\begin{align}
    R =R_{0} + \frac{\bar{R}}{1-P^{2}\left(\cos\theta_{\mathrm{eq}}m_{z'} - \sin\theta_{\mathrm{eq}}m_{y'}\right)}.
\end{align}
Up to the first order of $m_{y'}$, the resistance can be approximated as
\begin{align}
    \label{eq:Rmy}
    R =R_{0} + \frac{\bar{R}}{1-P^{2}\cos\theta_{\mathrm{eq}}}
    -
    \frac{\bar{R} P^{2}\sin\theta_{\mathrm{eq}}}{\left(1-P^{2}\cos\theta_{\mathrm{eq}}\right)^{2}}
    m_{y'}.
\end{align}

\section{Results}
In this section, we show the results of our theoretical analysis on magnetic 1/$f$ noise of a tiny TMR sensor. We first show the relation between the voltage power spectrum and the power spectrum of $m_{y'}$ in Sec. \ref{sec:psv}. To calculate the power spectrum of $m_{y'}$, we solve the linearized equations of motion of $\bm{m}$ by using the Fourier transformation in Sec. \ref{sec:eom}. Then we derive the Lorentzian power spectrum of $m_{y'}$ in Sec. \ref{sec:smy}. Assuming that bath correlation time, $\tau_{c}$, has a wide distribution, we derive the 1/$f$ power spectrum of voltage by superimposing the Lorentzian power spectra with different $\tau_{c}$ in Sec. \ref{sec:ps}. In Sec. \ref{sec:comp}, we show that the Hooge parameter of a tiny TMR sensor is much smaller than the conventional TMR sensor with the same sensitivity by comparing with the experimental results of Ref. \cite{Nakatani2022}.

\subsection{Power spectrum of voltage}
\label{sec:psv}
In most experiments, the voltage noise of a TMR sensor is measured under a constant direct current, $I$. Assuming that the measured voltage, $V$, is proportional to the resistance, $R$, the power spectrum of voltag, $S_{VV}(f)$, is proportional to the power spectrum of resistance, $S_{RR}(f)$, as
\begin{align}
    \label{eq:SVVdef}
    S_{VV}(f)=I^{2}S_{RR}(f).
\end{align}
Introducing the angular frequency, $\omega=2\pi f$, the power spectrum of resistance is defined as
\begin{align}
    \label{eq:SRRdef}
    S_{RR}(\omega)
    =
    4\int_{0}^{\infty}
    \left\langle R(t)R(0)\right\rangle \cos(\omega t)  dt,
\end{align}
where $\langle\ \rangle$ represents the statistical average.
Substituting Eq. \eqref{eq:Rmy} into Eq. \eqref{eq:SRRdef}, $S_{RR}(\omega)$ is expressed as
\begin{align}
    \label{eq:SRR_omega}
    S_{RR}(\omega)
    =
    \left[
        \frac{\bar{R} P^{2}\sin\theta_{\mathrm{eq}}}{\left(1-P^{2}\cos\theta_{\mathrm{eq}}\right)^{2}}
        \right]^{2}
    S_{m_{y'}m_{y'}}(\omega),
\end{align}
where $S_{m_{y'}m_{y'}}(\omega)$ is the power spectrum of $m_{y'}$ defined as
\begin{align}
    \label{eq:Smymy_def}
    S_{m_{y'}m_{y'}}(\omega)
    =    4 \int_{0}^{\infty}
    \left\langle m_{y'}(t)m_{y'}(0)\right\rangle \cos(\omega t)  dt.
\end{align}

We define the Fourier transform of a function $f(t)$ as $f(\omega) = \int_{-\infty}^{\infty} f(t) \exp\left(-i \omega t\right)  dt$. Substituting the inverse Fourier transform of $m_{y'}$ into Eq. \eqref{eq:Smymy_def} and performing some algebra, we obtain
\begin{align}
    \label{eq:smymy}
    S_{m_{y'}m_{y'}}(\omega)
     & =
    \frac{1}{2\pi}
    \int_{-\infty}^{\infty}
    \langle m_{y'}(\omega) m_{y'}(\omega') \rangle
    d\omega'
    \nonumber \\
     &
    +
    \frac{1}{2\pi}
    \int_{-\infty}^{\infty}
    \langle m_{y'}(-\omega) m_{y'}(\omega') \rangle
    d\omega'.
\end{align}
The Fourier transform of $m_{y'}$ can be obtained by solving the equations of motion in the Fourier space.

\subsection{Equations of motion and the Fourier transforms of $m_{x}$ and $m_{y'}$}
\label{sec:eom}
The equations of motion of $\bm{m}$ is given by the following generalized Langevin equation \cite{Kawabata1972,Miyazaki1998a,Imamura2022},
\begin{align}
    \label{eq:eom_def}
    \dot{\bm{m}}(t)
     & =
    -\gamma \bm{m}(t)\times\left(\bm{H}_{\mathrm{eff}} + \bm{r}\right)
    \nonumber \\
     &
    +\alpha  \bm{m}\times\int_{-\infty}^{t}\nu(t-t')\dot{\bm{m}}(t')  dt',
\end{align}
where $\dot{\bm{m}}(t)$ is the time derivative of $\bm{m}(t)$, $\gamma$ is the gyromagnetic ratio, and $\alpha$ is the Gilbert damping constant. The effective magnetic field acting on $\bm{m}$ is given by
\begin{align}
    \bm{H}_{\mathrm{eff}}
    =
    -M_{s} m_{x} \bm{e}_{x}
    +H_{b} \bm{e}_{y}
    +\left(H_{p} +H_{k} m_{z}\right) \bm{e}_{z}.
\end{align}
The memory function is defined as
\begin{align}
    \nu(t-t') = \frac{1}{\tau_{c}}\exp\left(-\frac{\left|t-t'\right|}{\tau_{c}}\right),
\end{align}
where $\tau_{c}$ is the bath correlation time.
The thermal agitation field, $\bm{r}$, is a random field satisfying $\langle r_{j}\rangle = 0$ and
\begin{align}
    \label{eq:r_corr}
     & \langle r_{j}r_{k}\rangle = \frac{\mu}{2}\delta_{j,k}\nu(t-t'),
\end{align}
where subscripts $j$ and $k$ denotes $x$, $y$, $z$, $y'$, or $z'$. The constant $\mu$ is defined as
\begin{align}
    \label{eq:mu}
    \mu=\frac{2\alpha k_{B}T}{\gamma \mu_{0} M_{s} \Omega},
\end{align}
where $k_{B}$ is the Boltzmann constant, $T$ is temperature, and $\Omega$ is the volume of the FL. From Eqs. \eqref{eq:r_corr} and \eqref{eq:mu} we see that the magnitude of the thermal agitation field is of the order of $\sqrt{\alpha}$ because $\mu$ is of the order of $\alpha$. The stochastic LLG equation with the Markovian damping derived by Brown \cite{Brown1963} is reproduced in the limit of $\tau_{c}\to 0$ because $\lim_{\tau_{c}\to 0} \nu(t-t') = 2\delta(t-t')$, where $\delta(t-t')$ is Dirac's delta function. It should be noted that 1/$f$ noise cannot be derived from the LLG equation with the Markovian damping because many physical processes with different time scale is required to generate 1/$f$ noise.

Since the FL of a typical TMR sensor is made of a ferromagnetic material with $\alpha \ll 1$, we focus on terms up to the first order of $\alpha$ in the equations of motion. We also assume that $m_{x}$, $m_{y'}$, $r_{x}$, $r_{y'}$, and $r_{z'}$ are small enough to linearize the equations of motion in terms of these small variables. Equation \eqref{eq:eom_def} can be approximated as
\begin{align}
    \label{eq:eom_1}
     & \dot{m}_{x}(t) = -\omega_{0}  m_{y'}(t) +\gamma r_{y'}(t)
    \nonumber                                                    \\
     & \hspace{4em}
    -\alpha\int_{-\infty}^{t}\nu(t-t') \dot{m}_{y'}(t')  dt'     \\
    \label{eq:eom_2}
     & \dot{m}_{y'}(t) = \omega_{1}  m_{x}(t) -\gamma r_{x}(t)
    \nonumber                                                    \\
     & \hspace{4em}
    +\alpha\int_{-\infty}^{t}\nu(t-t') \dot{m}_{x}(t')  dt'      \\
     & \dot{m}_{z'}(t)=0,
\end{align}
where
\begin{align}
    \omega_{0}
     & =
    \gamma
    \left(
    H_{b}\sin\theta_{\rm eq}
    + H_{p}\cos\theta_{\rm eq}
    + H_{k}\cos2\theta_{\rm eq}
    \right),
    \\
    \omega_{1}
     & =
    \gamma\left(
    M_{s}
    \! + \! H_{b} \sin\theta_{\rm eq}
    \! + \! H_{p} \cos\theta_{\rm eq}
    \! + \! H_{k}\cos^{2}\theta_{\rm eq}
    \right).
\end{align}
Following Ref. \cite{Imamura2022}, we approximate the non-Markovian damping term in Eqs. \eqref{eq:eom_1} and  \eqref{eq:eom_2} up to the first order of $\alpha$. Successive application of the integration by parts gives the following linearized equations of motion up to the order of $\alpha$,
\begin{align}
    \label{eq:eom_x}
     & \dot{m}_{x}(t)
    =
    -\hat{\gamma}_{1}\omega_{0}  m_{y'}(t)
    +\gamma r_{y'}(t)
    -\tilde{\alpha}\omega_{1}  m_{x}(t)
    \\
    \label{eq:eom_y}
     & \dot{m}_{y'}(t)
    =
    \hat{\gamma}_{0}\omega_{1}  m_{x}(t)
    - \gamma  r_{x}(t)
    -\tilde{\alpha}\omega_{0}  m_{y'}(t),
\end{align}
where
\begin{align}
    \label{eq:hat_gamma0}
     & \hat{\gamma}_{0} = \left(1+ \frac{\alpha \xi_{0}}{1+\xi_{0}\xi_{1}}\right) \\
    \label{eq:hat_gamma1}
     & \hat{\gamma}_{1} = \left(1+ \frac{\alpha \xi_{1}}{1+\xi_{0}\xi_{1}}\right) \\
    \label{eq:tilde_alpha}
     & \tilde{\alpha} =  \frac{\alpha}{1+\xi_{0}\xi_{1}}                          \\
     & \xi_{0} = \tau_{c} \omega_{0}\nonumber                                     \\
     & \xi_{1} = \tau_{c} \omega_{1}.
\end{align}
Details of the derivation of the above equations will be provided in Appendix \ref{sec:aa}.
In the Fourier space, the equations of motion are expressed as
\begin{align}
     & i\omega  m_{x}(\omega)
    =
    -\hat{\gamma}_{1}\omega_{0} m_{y'}(\omega)
    +\gamma r_{y'}(\omega)
    -\tilde{\alpha} \omega_{1} m_{x}(\omega),
    \\
     & i\omega  m_{y'}(\omega)
    =
    \hat{\gamma}_{0} \omega_{1}  m_{x}(\omega)
    - \gamma  r_{x}(\omega)
    -\tilde{\alpha}\omega_{0}  m_{y'}(\omega).
\end{align}
The solutions are obtained as
\begin{align}
    \label{eq:mxomega}
     & m_{x}(\omega)
    =
    \frac{ \hat{\gamma}_{1}\omega_{0}  \gamma  r_{x}(\omega)
        + \left(\tilde{\alpha}\omega_{0} + i\omega\right)\gamma  r_{y'}(\omega)}{A(\omega)}
    \\
    \label{eq:myomega}
     & m_{y'}(\omega)
    =
    \frac{\hat{\gamma}_{0} \omega_{1} \gamma r_{y'}(\omega)
        -\left(\tilde{\alpha}\omega_{1}+i\omega\right)\gamma r_{x}(\omega)}{A(\omega)},
\end{align}
where
\begin{align}
    A(\omega)
    =
    \left(\hat{\gamma}_{0}\hat{\gamma}_{1}+\tilde{\alpha}^{2}\right)
    \omega_{0}\omega_{1} - \omega^{2}
    + i\tilde{\alpha}(\omega_{0} + \omega_{1})\omega.
\end{align}

\subsection{Power spectrum of $m_{y'}$}
\label{sec:smy}

From Eq. \eqref{eq:myomega}, the correlation of $m_{y'}(\omega)$ and $m_{y'}(\omega')$ is expressed as
\begin{align}
    \label{eq:my_cor1}
     & \langle m_{y'}(\omega) m_{y'}(\omega') \rangle
    =
    \frac{\left(\hat{\gamma}_{0}\omega_{1}\right)^{2}}{A(\omega)A(\omega')}
    \gamma^{2} \left\langle r_{y'}(\omega) r_{y'}(\omega')
    \right\rangle
    \nonumber
    \\
     &
    +\frac{
        \left(\tilde{\alpha}\omega_{1}+i\omega\right)
        \left(\tilde{\alpha}\omega_{1}+i\omega'\right)
    }{A(\omega)A(\omega')}
    \gamma^{2}   \left\langle r_{x}(\omega) r_{x}(\omega')    \right\rangle,
\end{align}
where we use the fact that $r_x$ and $r_{y'}$ do not correlate with each other. The correlation $\langle m_{y'}(-\omega) m_{y'}(\omega')\rangle$ is obtained by replacing $\omega$ with $-\omega$ in Eq. \eqref{eq:my_cor1}

Following Ref. \cite{Kubo1966}, the correlation of thermal agitation fields in the Fourier space is obtained as
\begin{align}
    \label{eq:r_cor1}
    \langle r_{j}(\omega)r_{k}(\omega')\rangle
    = 2\pi\mu\delta_{j,k}\frac{1}{1+i\tau_{c}\omega}\delta(\omega+\omega').
\end{align}
The correlation $\langle r_{j}(-\omega)r_{k}(\omega')\rangle$ is obtained by replacing $\omega$ with $-\omega$ in Eq. \eqref{eq:r_cor1}.

Substituting Eqs. \eqref{eq:my_cor1} and \eqref{eq:r_cor1} into Eq. \eqref{eq:smymy}, the power spectrum of $m_{y'}$ is expressed as
\begin{align}
    \label{eq:syy3}
    S_{m_{y'}m_{y'}}(\omega)
     & =
    \frac{  \left(
        \hat{\gamma}_{0}^{2} + \tilde{\alpha}^{2}\right) \omega_{1}^{2}
        +\omega^{2}}{B(\omega)}
    \frac{2\gamma^{2}\mu}{1+(\tau_{c}\omega)^{2}},
\end{align}
where
\begin{align}
    B(\omega)
     & =
    \left[
        \left(\hat{\gamma}_{0}\hat{\gamma}_{1}+\tilde{\alpha}^{2}\right)
        \omega_{0}\omega_{1} - \omega^{2}
        \right]^{2}
    \nonumber \\
     &
    \hspace{1em}
    +\left[\tilde{\alpha}(\omega_{0} + \omega_{1})\omega\right]^{2}.
\end{align}

In the low frequency regime satisfying $\omega \ll \omega_{0}$ and $\omega\ll\omega_{1}$,  Eq. \eqref{eq:syy3} can be approximated by the Lorentzian function as
\begin{align}
    \label{eq:syy4}
    S_{m_{y'}m_{y'}}(\omega)
     & =
    \frac{
        2\gamma^{2}\mu
    }{
        \left(
        \hat{\gamma}_{1}\omega_{0}
        \right)^{2}
    }
    \frac{1}{1+(\tau_{c}\omega)^{2}}.
\end{align}
Since $\omega_{0}$ and $\omega_{1}$ are of the order of 0.1 GHz $\sim$ 10 GHz for conventional TMR sensors \cite{Nakatani2022}, the low frequency condition is clearly satisfied for the frequency range of the bio-magnetic signal, i.e. less than a few hundred Hz.

\subsection{Superimposition of Lorentzian power spectra}
\label{sec:ps}
Bath correlation time, $\tau_{c}$, is the decay time of the correlation of thermal agitation field as shown in Eq. \eqref{eq:r_corr}. Thermal agitation field is produced by many kinds of sources or baths such as dipolar coupling with magnons in the reference layer and spin orbit coupling with phonons. Since $\tau_{c}$ depends on the relaxation mechanism of the bath, different relaxation modes in different baths have their own $\tau_{c}$. Instead of discussing $\tau_{c}$ for some specific types of baths, we just assume a distribution of $\tau_{c}$ and analyze the effect of the distribution of $\tau_{c}$ on the low frequency power spectrum of voltage. Assuming a wide distribution of $\tau_{c}$, we derive an analytical expression of the  power spectrum of the magnetic 1/$f$ noise.

We assume that $\tau_{c}$ is uniformly distributed in the range of $\tau_{c,\mathrm{min}}\le \tau_{c} \le\tau_{c,\mathrm{max}}$ and has the probability distribution defined as $\rho(\tau_{c})=1/(\tau_{c,\mathrm{max}}-\tau_{c,\mathrm{min}})$.
The superimposition of $S_{m_{y'}m_{y'}}(\omega)$ for all $\tau_{c}$ is given by
\begin{align}
    \label{eq:syy5}
    S_{m_{y'}m_{y'}}(\omega)
     & =
    \frac{2\gamma^{2}\mu}{\omega_{0}^{2}}
    \int_{0}^{\infty}
    \frac{1}{\hat{\gamma}_{1}^{2}}
    \frac{\rho(\tau_{c})}{1+(\tau_{c}\omega)^{2}}
    d\tau_{c}.
\end{align}
As a function of $\tau_c$, $\hat{\gamma}_{1}$ is almost unity except around the peak at $\tau_{c} = 1/\sqrt{\omega_{0}\omega_{1}}$, which is of the order of ns. Since the peak value of $\hat{\gamma}_{1}$ is as small as $1+(\alpha/2)\sqrt{\omega_{1}/\omega_{0}}$, and $\omega$ is assumed to be much smaller than $\omega_{0}$ and $\omega_{1}$, Eq. \eqref{eq:syy5} can be approximated as
\begin{align}
    \label{eq:syy8}
    S_{m_{y'}m_{y'}}(\omega)
     & =
    \frac{2\gamma^{2}\mu}{\omega_{0}^{2}}
    \frac{1}{\tau_{c,\mathrm{diff}}}
    \int_{\tau_{c,\mathrm{min}}}^{\tau_{c,\mathrm{max}}}
    \frac{1}{1+(\tau_{c}\omega)^{2}}
    d\tau_{c}
    \nonumber       \\
     & =
    \frac{2\gamma^{2}\mu}{\omega_{0}^{2}}
    \frac{1}{\tau_{c,\mathrm{diff}}}
    \left[\frac{\arctan\left(\omega\tau_{c,\mathrm{max}}\right)}{\omega}\right.
    \nonumber       \\
     & \hspace{1em}
        \left.
        -
        \frac{\arctan\left(\omega\tau_{c,\mathrm{min}}\right)}{\omega}
        \right],
\end{align}
where $\tau_{c,\mathrm{diff}} = \tau_{c,\mathrm{max}} - \tau_{c,\mathrm{min}}$.
Since $\arctan(x)/x$ is a monotonically decreasing function of $x$ for $x>0$ and $\lim_{x\to 0} \arctan(x)/x = 1$, $S_{m_{y'}m_{y'}}(\omega)$ is a monotonically decreasing function of $\omega$ and take a maximum value of $2\gamma^{2}\mu/\omega_{0}^{2}$ in the limit of $\omega\to 0$.

\begin{figure}[t]
    \centerline{ \includegraphics[width=0.9\columnwidth]{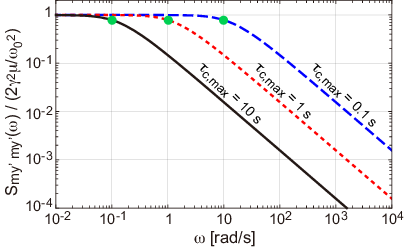}}
    \caption{
    \label{fig:fig2}
    Power spectrum of $m_{y'}$, $S_{m_{y'}m_{y'}}(\omega)$ given by Eq. \eqref{eq:atan} normalized by $2\gamma^{2}\mu/\omega_{0}^{2}$. The black solid, red dotted, and blue dashed curves represent the results for $\tau_{c,\mathrm{max}}$=10, 1, and 0.1 s, respectively. The green circles indicate the values at $\omega$ = $1/\tau_{c,\mathrm{max}}$.
    }
\end{figure}

When $\omega\tau_{c,\mathrm{min}}\ll 1$, the second term in the square bracket of Eq. \eqref{eq:syy8} can be neglected and $S_{m_{y'}m_{y'}}(\omega)$ is approximated as
\begin{align}
    \label{eq:atan}
    S_{m_{y'}m_{y'}}(\omega)
    =
    \frac{2\gamma^{2}\mu}{\omega_{0}^{2}}
    \frac{\arctan\left(\omega\tau_{c,\mathrm{max}}\right)}{\omega\tau_{c,\mathrm{max}}}.
\end{align}
Figure \ref{fig:fig2} shows $S_{m_{y'}m_{y'}}(\omega)$ given by Eq. \eqref{eq:atan} normalized by $2\gamma^{2}\mu/\omega_{0}^{2}$ for $\tau_{c,\mathrm{max}}$=10 s (black solid), 1 s (red dotted), and 0.1 s (blue dashed). The values at $\omega$ = $1/\tau_{c,\mathrm{max}}$ are indicated by the green circles. All curves are almost flat for $\omega \ll 1/\tau_{c,\mathrm{max}}$ and inversely proportional to $\omega$ for $\omega \gg 1/\tau_{c,\mathrm{max}}$.

Assuming a wide distribution of $\tau_{c}$ satisfying $\omega\tau_{c,\mathrm{min}}\ll 1$ and $\omega\tau_{c,\mathrm{max}}\gg 1$, we have $\arctan\left(\omega\tau_{c,\mathrm{min}}\right)= 0$ and $\arctan\left(\omega\tau_{c,\mathrm{max}}\right)= \pi/2$. Then the power spectrum can be approximated as
\begin{align}
    \label{eq:syy9}
    S_{m_{y'}m_{y'}}(\omega)
     & =
    \frac{2\gamma^{2}\mu}{\omega_{0}^{2}}
    \frac{1}{\tau_{c,\mathrm{max}}}
    \frac{\pi}{2\omega},
\end{align}
which is inversely proportional to the angular frequency, $\omega$ (=$2\pi f$). From Eqs. \eqref{eq:SVVdef}, \eqref{eq:SRR_omega} and \eqref{eq:syy9} the voltage power spectrum is given by
\begin{align}
    \label{eq:SRR1}
    S_{VV}(f)
     & =
    \left[
        \frac{I\bar{R} P^{2}\sin\theta_{\mathrm{eq}}}{\left(1-P^{2}\cos\theta_{\mathrm{eq}}\right)^{2}}
        \right]^{2}
    \frac{\gamma^{2}\mu}{2\omega_{0}^{2}}
    \frac{1}{\tau_{c,\mathrm{max}}}
    \frac{1}{f}.
\end{align}
{\em This is the main result of this paper}. The obvious difference from  other models of low frequency magnetic noise \cite{Hardner1993,Ingvarsson2000,Smith2001,Safonov2002,Jiang2004,Ozbay2009} is that Eq. \eqref{eq:SRR1} has the term $1/\tau_{c,\mathrm{max}}$ as information of the distribution of the bath correlation time.
It should be noted that the 1/f noise of a tiny TMR sensor we derived is response to the thermal agitation fields that exhibit 1/f power spectrum as the superimposition of the Lorentzian power spectrum.

\subsection{
    Comparison with a conventional TMR sensor with large junction area}
\label{sec:comp}
We compare the derived 1/$f$ noise of the macrospin model with the experimental results of a conventional TMR sensor with large junction area reported in Ref. \cite{Nakatani2022}. The Hooge parameter, $\alpha_{H}$, is a convenient measure to compare the 1/f noise between different MTJs, which is defined as
\begin{align}
    \label{eq:Hooge}
    S_{VV}(f) = S_{VV}^{\rm wh} + \alpha_{H}V_{b}^{2} A^{-1} f^{-1},
\end{align}
where $S_{VV}^{\rm wh}$ is the power spectral density of the white noise, $V_{b}$ is the bias voltage, and $A$ is the area of the MTJ. The typical value of the Hooge parameter of conventional TMR sensors is about $10^{-6} \sim 10^{-11}$ $\mu$m$^{2}$ \cite{Silva2015,Garcia2021,Nakatani2022,Matos2023}.
From Eq. \eqref{eq:Rmy}, the bias voltage is given by
\begin{align}
    \label{eq:vb}
    V_{b}=I
    \left(
    R_{0} + \frac{\bar{R}}{1-P^{2}\cos\theta_{\mathrm{eq}}}
    \right).
\end{align}
From Eqs. \eqref{eq:mu}, \eqref{eq:SRR1}, \eqref{eq:Hooge}, and  \eqref{eq:vb}, the Hooge parameter of a tiny TMR sensor is obtained as
\begin{align}
    \label{eq:ah}
    \alpha_{H}
     & =\left\{
    \frac{\bar{R}P^{2}\sin\theta_{\rm eq}}{(1-P^{2}\cos\theta_{\rm eq})\left[\bar{R} + R_{0}(1-P^{2}\cos\theta_{\rm eq})\right]}
    \right\}^{2}
    \nonumber                                                                                                                    \\
     & \times \frac{\alpha\, \gamma\, k_{\rm B} T}{\mu_{0}\, M_{s}\, d}\frac{1}{\omega_{0}^{2}}\frac{1}{\tau_{c, \mathrm{max}}},
\end{align}
where $d$ is the thickness of the FL.

\begin{figure}[t]
    \centerline{ \includegraphics[width=\columnwidth]{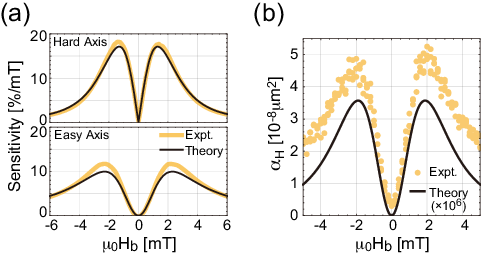}}
    \caption{
    \label{fig:fig3}
    (a) Sensitivity as a function of the bias field, $\mu_{0}H_{b}$. The top panel shows the results for the signal field along the magnetization hard axis, i.e the $y$ axis. The bottom panel shows the results for the signal field along the magnetization easy axis, i.e $z$ axis. In both panels, the yellow and the black curves represent the experimental and theoretical results, respectively.
    (b) Hooge parameter, $\alpha_{H}$, as a function of the bias field, $\mu_{0}H_{b}$. The yellow circles and the black curve represent the experimental and theoretical results, respectively. Note that the theoretical results are multiplied by $10^{6}$. In all panels, the experimental results are the same as those shown in Fig. 3(b) in Ref. \cite{Nakatani2022}.
    }
\end{figure}

To compare Eq. \eqref{eq:ah} with the experimental results of a conventional TMR sensor, we determine the junction parameters by fitting the bias field dependence of the resistance shown in Figs. 2(a) of Ref. \cite{Nakatani2022}. The parameters are determined as $M_{s}$=0.93 MA/m, $\mu_{0}H_{k}$=1.0 mT, $\mu_{0}H_{p}$= 2.15 mT, $R_{0}$=10.7 $\Omega$, $\bar{R}$=10.3 $\Omega$, $P^{2}$=0.74. Figure \ref{fig:fig3}(a) shows the bias field, $\mu_{0}H_{b}$, dependence of the sensitivity defined as
\begin{align}
    {\rm Sensitivity} = \frac{1}{R_{\rm max}}\frac{d R}{d\, \mu_{0}H_{b}},
\end{align}
where $R_{\rm max}$ is the maximum value of the resistance. The experimental results indicated by the yellow curves are well reproduced by the theoretical results represented by the black curves.

Figure \ref{fig:fig3}(b) shows the bias field dependence of the Hooge parameter, $\alpha_{H}$.
The experimental results are indicated by the yellow circles. The black solid curve represents the theoretical results multiplied by $10^6$. For calculation of the Hooge parameter, the following parameters are assumed: $\alpha$ = 0.05, $d$ = 2 nm, $\tau_{c, \mathrm{max}}$ = 1 s, and $T$ = 300 K. The field dependences of the Hooge parameter and the sensitivity look very similar because the Hooge parameter is proportional to the square of the sensitivity along the $y'$ axis as indicated by Eqs. (6), (42), and (46).

The results show that the Hooge parameter of a tiny TMR sensor is much smaller than that of the conventional TMR sensor with the same sensitivity by a factor of $10^{-6}$. If a number of tiny MTJs are connected in parallel to reproduce the same resistance as a conventional TMR sensor, the power spectrum of the magnetic 1/$f$ noise can be reduced by a factor of $10^{-6}$ without reducing sensitivity.

\section{Summary}
In summary, we propose a theoretical model for magnetic 1/$f$ noise of a tiny TMR sensor originated from distribution of bath correlation time. Starting from the generalized Langevin equation, we derive an analytical expression of the low frequency power spectrum of voltage. Assuming a wide distribution of the bath correlation times, the derived voltage power spectrum is inversely proportional to the frequency. We also show that the Hooge parameter of a tiny TMR sensor is much smaller than that of a conventional TMR sensor with large junction area. The power spectrum of the 1/$f$ noise can be substantially reduced without reducing sensitivity by connecting tiny TMR sensors in parallel. The results provides a new perspective on magnetic 1/$f$ noise and will be useful for reduction of 1/$f$ noise of TMR sensors. The presented theoretical framework is applicable not only to the magnetic 1/$f$ noise of a tiny TMR sensor, but also to the low frequency fluctuation of any tiny magnetic devices where the macrospin model is appropriate.

\begin{acknowledgments}
    The authors thank T. Nakatani for providing experimental data and for valuable discussions. This work was partly supported by JSPS KAKENHI Grant No. JP23K04575.
\end{acknowledgments}

\appendix
\section{Derivation of Eqs. \eqref{eq:eom_x} and \eqref{eq:eom_y} }
\label{sec:aa}
In this section, we provide the details of the derivation of  Eqs. \eqref{eq:eom_x} and \eqref{eq:eom_y}.
Following Ref. \cite{Imamura2022}, different approaches are used depending on the value of $\tau_{c}$: in the short $\tau_{c}$ regime and in the long $\tau_{c}$ regime. Introducing $\xi_{0} = \tau_{c} \omega_{0}$ and $\xi_{1} = \tau_{c} \omega_{1}$, the short $\tau_{c}$ regime is defined as $\xi_{0}\xi_{1}<1$, and the long $\tau_{c}$ regime is defined as $\xi_{0}\xi_{1}>1$. For both $\tau_{c}$ regimes we can derive the same equations of motion as Eqs. \eqref{eq:eom_x} and \eqref{eq:eom_y}.
\subsubsection{Short $\tau_{c}$ regime:  $\xi_{0}\xi_{1}<1$ }
We approximate the non-Markovian damping term in
Eq. \eqref{eq:eom_1} up to the first order of $\alpha$. Successive application of the integration by parts
using $ \nu(t-t')=\tau_{c} \left[d\nu(t-t')/dt'\right]$ the integral part of the non-Markovian damping term in Eq. \eqref{eq:eom_1} is expressed as
\begin{align}
    \label{eq:nmint_def1}
    \int_{-\infty}^{t}\nu(t-t') \dot{m}_{y'}(t')  dt'
    =
    \sum_{n=1}^{\infty} (-\tau_{c})^{n-1}\frac{d^{n}}{dt^{n}} m_{y'}(t).
\end{align}
Since the integral part of the non-Markovian damping is multiplied by $\alpha$ we approximate the time derivative of $m_{y'}(t)$ in the 0th order of $\alpha$ as
\begin{align}
    \label{eq:dmdt_even}
    \frac{d^{2n}}{dt^{2n}}m_{y'}(t)
    =
    (-1)^{n}\omega_{0}^{n} \omega_{1}^{n}  m_{y'}(t),
\end{align}
and
\begin{align}
    \label{eq:dmdt_odd}
    \frac{d^{2n+1}}{dt^{2n+1}}m_{y'}(t)
    =
    (-1)^{n}\omega_{0}^{n}\omega_{1}^{n+1}   m_{x}(t).
\end{align}
Substituting Eqs. \eqref{eq:dmdt_even} and \eqref{eq:dmdt_odd} into Eq. \eqref{eq:nmint_def1}, the integral part of the non-Markovian damping term in Eq. \eqref{eq:eom_1} is expressed as
\begin{align}
    \label{eq:nmint_1}
     & \int_{-\infty}^{t}\nu(t-t') \dot{m}_{y'}(t')  dt'
    =
    \Bigl[
        \xi_{1}  \omega_{0} m_{y'}(t)
    \nonumber                                            \\
     &
        +
        \omega_{1} m_{x}(t)
        \Bigr]\sum_{n=1}^{\infty} (-\xi_{0}\xi_{1})^{n-1}.
\end{align}
The summation in Eq. \eqref{eq:nmint_1} converges under the condition of
$\xi_{0}\xi_{1}<1$ as
\begin{align}
    \sum_{n=1}^{\infty}(-\xi_{0}\xi_{1})^{n-1} = \frac{1}{1+\xi_{0}\xi_{1}}.
\end{align}
Then Eq. \eqref{eq:nmint_1} becomes
\begin{align}
    \label{eq:nmint_a}
    \int_{-\infty}^{t}\nu(t-t') \dot{m}_{y'}(t')  dt'
    =
    \frac{\xi_{1}\omega_{0}m_{y'}(t) +\omega_{1} m_{x}(t)}{1+\xi_{0}\xi_{1}}.
\end{align}
Substituting Eq. \eqref{eq:nmint_a} into Eq. \eqref{eq:eom_1} and performing some algebra, we obtain the following linearized equation of motion for $m_{x}(t)$ up to the first order of $\alpha$:
\begin{align}
    \label{eq:eom_11}
    \dot{m}_{x}(t)
     & =
    -\left(1+ \frac{\alpha \xi_{1}}{1+\xi_{0}\xi_{1}}\right)\omega_{0}  m_{y'}(t)
    \nonumber \\
     &
    +\gamma r_{y'}(t)
    -\frac{\alpha}{1+\xi_{0}\xi_{1}}\omega_{1}  m_{x}(t).
\end{align}

Similarly the following linearized equation of motion for $m_{y'}(t)$ up to the first order of $\alpha$ is obtained as
\begin{align}
    \label{eq:eom_21}
    \dot{m}_{y'}(t)
     & =
    \left(1+ \frac{\alpha \xi_{0}}{1+\xi_{0}\xi_{1}}\right)\omega_{1}  m_{x}(t)
    \nonumber \\
     &
    -\gamma r_{x}(t)
    -\frac{\alpha}{1+\xi_{0}\xi_{1}}\omega_{0}  m_{y'}(t).
\end{align}
Using the symbols defined by Eqs. \eqref{eq:hat_gamma0}, \eqref{eq:hat_gamma1}, and \eqref{eq:tilde_alpha}, one can easily confirm that Eqs. \eqref{eq:eom_11} and \eqref{eq:eom_21} are the same as Eqs. \eqref{eq:eom_x} and \eqref{eq:eom_y}, respectively.

\subsubsection{Long $\tau_{c}$ regime: $\xi_{0}\xi_{1}>1$}
In the long bath $\tau_{c}$ regime satisfying $\xi_{0}\xi_{1} > 1$, we expand Eq. \eqref{eq:eom_1} in power series of $1/(\xi_{0}\xi_{1})$. Using the integration by parts with $d\nu(t-t')/dt'=\nu(t-t')/\tau_{c}$
the integral part of the non-Markovian damping in Eq. \eqref{eq:eom_1} can be written as
\begin{align}
     & \int_{-\infty}^{t}
    \nu(t-t')  \dot{m}_{y'}(t')  dt'
    =
    \frac{1}{\tau_{c}}
    \int_{-\infty}^{t}\dot{m}_{y'}(t')  dt'
    \nonumber             \\
     &
    -\frac{1}{\tau_{c}}
    \int_{-\infty}^{t}
    \nu(t-t')
    \left[
        \int_{-\infty}^{t'}
        \dot{m}_{y'}(t'')  dt''
        \right]
    dt'.
\end{align}
Successive application of the integration by parts gives
\begin{align}
    \label{eq:nmint_def_3}
    \int_{-\infty}^{t}\nu(t-t') \dot{m}_{y'}(t')  dt'
    =
    -\sum_{n=1}^{\infty}
    \left(-\frac{1}{\tau_{c}}\right)^{n}
    J_{n},
\end{align}
where $J_{n}$ is the $n$th order multiple integral defined as
\begin{align}
    \label{eq:Jn_def}
    J_{n}=\int_{-\infty}^{t}\int_{-\infty}^{t_{1}}\cdots\int_{-\infty}^{t_{n-1}}
    \dot{m}_{y'}(t_{n })  dt_{n}\cdots dt_{2} dt_{1}.
\end{align}
From Eq. \eqref{eq:dmdt_even}, on the other hand, $\dot{m}_{y'}(t)$ is expressed as
\begin{align}
    \label{eq:mydot}
    \dot{m}_{y'}(t)
    =
    \frac{1}{(-1)^{n} \omega_{0}^{n} \omega_{1}^{n}}
    \left(
    \frac{d^{2n+1}}{dt^{2n+1}}m_{y'}(t)
    \right).
\end{align}
Substituting Eq. \eqref{eq:mydot} into Eq. \eqref{eq:Jn_def} the multiple integrals are calculated as
\begin{align}
    \label{eq:j2n0}
    J_{2n}
     & =
    \frac{1}{(-1)^{n} \omega_{0}^{n} \omega_{1}^{n}}
    \dot{m}_{y'}(t),
\end{align}
and
\begin{align}
    \label{eq:j2nm10}
    J_{2n-1}
     & =
    \frac{1}{(-1)^{n} \omega_{0}^{n} \omega_{1}^{n}}
    \ddot{m}_{y'}(t).
\end{align}
Substituting the time derivative of $m_{y'}(t)$ in the 0th order of $\alpha$ into Eqs. \eqref{eq:j2n0} and \eqref{eq:j2nm10}, we obtain
\begin{align}
    \label{eq:j2n}
    J_{2n}
    =
    \frac{\omega_{1}m_{x}(t)}{(-1)^{n}\omega_{0}^{n} \omega_{1}^{n}},
\end{align}
and
\begin{align}
    \label{eq:j2nm1}
    J_{2n-1}
    =
    \frac{m_{y'}(t)}{(-1)^{n-1}\omega_{0}^{n-1} \omega_{1}^{n-1}}.
\end{align}
Substituting Eqs. \eqref{eq:j2n} and \eqref{eq:j2nm1} into Eq. \eqref{eq:nmint_def_3} and performing some algebra, the integral part of the non-Markovian damping in Eq. \eqref{eq:eom_1} can be expressed as
\begin{align}
    \label{eq:nmint_3}
     &
    \int_{-\infty}^{t}\nu(t-t') \dot{m}_{y'}(t')  dt'\nonumber \\
     & =
    -\sum_{n=1}^{\infty}
    \left[
    \left(-\frac{1}{\tau_{c}}\right)^{2n-1}J_{2n-1} + \left(-\frac{1}{\tau_{c}}\right)^{2n}J_{2n}
    \right]
    \nonumber                                                  \\
     & =
    -
    \left[
        \sum_{n=1}^{\infty}
        \left(-\frac{1}{\xi_{0} \xi_{1}}\right)^{n}
        \right]
    \Bigl[
        \xi_{1} \omega_{0} m_{y'}(t)
        +
        \omega_{1} m_{x}(t)
        \Bigr]
    \nonumber                                                  \\
     & =
    \frac{1}{1+\xi_{0}\xi_{1}}\left[\xi_{1} \omega_{0} m_{y'}(t)
    +
    \omega_{1}m_{x}(t)\right].
\end{align}
In the last equality, we use the following relation:
\begin{align}
    \sum_{n=1}^{\infty}
    \left(-\frac{1}{\xi_{0} \xi_{1}}\right)^{n}
    =
    -\frac{1}{1+\xi_{0}\xi_{1}},
\end{align}
which holds under the condition that $\xi_{0}\xi_{1}>1$. Equation \eqref{eq:nmint_3} is the same as Eq. \eqref{eq:nmint_a}. Substituting Eq. \eqref{eq:nmint_3} into Eq. \eqref{eq:eom_1} and preforming some algebra, we obtain the same linearized equation of motion of $m_{x}(t)$ as Eq. \eqref{eq:eom_x}. Equation \eqref{eq:eom_y} can also be obtained by similar calculations.

%

\end{document}